\documentclass[11pt,a4paper]{article}
\usepackage[noadjust]{cite}

\usepackage[margin=2.8cm,bottom=3.5cm]{geometry}

\usepackage{amsmath, amssymb}
\usepackage{color}
\pdfoutput=1 

\usepackage{graphicx} 
\usepackage{subcaption}
\usepackage[T1]{fontenc}

\title{Interaction between kinks and antikinks with double long-range tails}

\author{
        Jo\~ao G. F. Campos \\
        Departamento de F\'isica, Universidade Federal de Pernambuco,\\
        Av. Prof. Moraes Rego, 1235, Recife - PE - 50670-901, Brazil\\
        jgfc@df.ufpe.br
            \and
        Azadeh Mohammadi\\
        Departamento de F\'isica, Universidade Federal de Pernambuco,\\
        Av. Prof. Moraes Rego, 1235, Recife - PE - 50670-901, Brazil\\
        azadeh.mohammadi@ufpe.br
}

\begin{document} 
\maketitle

\begin{abstract}
We explore a class of $\phi^{4n}$ models with kink and antikink solutions that have long-range tails on both sides, specializing to the cases with $n=2$ and $n=3$. A recently developed method of an accelerating kink ansatz is used to estimate the force between the kink and the antikink. We use state-of-the-art numerical methods to initialize the system in a kink-antikink configuration with a finite initial velocity and to evolve the system according to the equations of motion. Among these methods, we propose a computationally efficient way to initialize the velocity field of the system. Interestingly, we discover that, for this class of models, $\phi^{4n}$ with $n>1$, the kink-antikink annihilation behaves differently from the archetypal $\phi^4$ model or even the kinks with one long-range tail because there is neither long-lived bion formation nor resonance windows and the critical velocity is ultrarelativistic.
\end{abstract}

\section{Introduction}

Soliton configurations in classical and quantum field theories are important solutions of field equations with applications in many areas of physics, ranging from condensed matter to cosmology 
\cite{manton2004topological,vilenkin2000cosmic}. In $(1+1)$ dimensions, a soliton configuration that interpolates between different minima is called a kink. The interaction between kinks or a kink and an antikink is an active area of research with a growing body of exciting work as, for instance, the study of kink-antikink creation from particles or radiation \cite{romanczukiewicz2006creation, dutta2008creating, romanczukiewicz2010oscillon}, multikink collisions \cite{marjaneh2017multi,marjaneh2017high,marjaneh2018extreme}, and the investigation of the role of quasinormal modes in kink-antikink collisions \cite{dorey2018resonant,campos2020quasinormal}. Notably, the study of kinks with quasinormal modes is interesting because the vibrational modes of the kink are important for the resonance phenomenon and fractal structure in kink-antikink interactions and turning the vibrational modes into quasinormal modes affect them. In particular, the work 
\cite{campos2020quasinormal} was our first contribution to this question.

Campbell et al. authored one of the seminal works in kink-antikink scattering \cite{campbell1983resonance}. In this work, the authors observed resonance windows in the kink-antikink collisions of the $\phi^4$ model, with the kink-antikink bouncing multiple times before separation. The authors provided an approximate explanation for this phenomenon as an energy exchange mechanism between the translational and vibrational modes of the kink. However, the analysis in \cite{campbell1983resonance} had a typo that was only corrected much later in \cite{takyi2016collective}, showing that this approximate model is not as effective to describe the resonance phenomenon as it seemed. 
One of the most intriguing properties of the resonance structure of kink-antikink collisions is that it exhibits a fractal structure, where a sequence of three-bounce windows exists adjacent to the two-bounce windows and similarly for higher-order resonance windows \cite{campbell1986kink,anninos1991fractal}. 
Curiously resonance windows were shown to also occur in the $\phi^6$ theory, despite the absence of vibrational modes in the single kink. In this case, the resonances arise from the interplay between the translational modes of the kink and antikink and a vibrational mode present in the kink-antikink pair configuration 
\cite{dorey2011kink}. 

Another intriguing question recently discussed in the literature is the investigation of soliton solutions in the presence of an impurity that preserves half of the BPS property of the system 
\cite{adam2019phi, adam2019spectral, adam2019bps, adam2019solvable, campos2020fermion}. The BPS property is relevant because it guarantees the stability of the system. Furthermore, adding an impurity that partially preserves the BPS property allows the study of BPS and non-BPS interactions between defects within the same model even in $(1+1)$ dimensions, where BPS interactions are absent otherwise. 

Other works worth mentioning, along the same line as before, include the collision of topological defects of two-component scalar fields 
\cite{halavanau2012resonance, alonso2018reflection}, application of the collective coordinates method to the $\phi^6$ model \cite{gani2014kink}, and many others \cite{forgacs2008negative,simas2017degenerate, bazeia2018scattering}.

The interaction between kinks with long-range tails is a highly nontrivial problem that has recently drawn some attention 
\cite{christov2019long, manton2019forces, khare2019family, christov2019kink,christov2020kink}. These tails decay as a power-law instead of the usual exponential decay due to higher-order terms in the Lagrangian potential. The power-law decay means that the overlap between the kinks is much more substantial and the kinks interaction is much stronger than the ones with exponential tail, making them fascinating for further studies. Besides this, field theories in higher dimensions and with higher-order polynomial potentials giving rise to long-range interacting solitons are also appealing due to their applications in cosmology \cite{greenwood2009electroweak, valle2016relativistic}, statistical mechanics \cite{campa2009statistical}, condensed matter \cite{saffman2010quantum, lahaye2009physics}, and supersymmetric quantum mechanics \cite{bazeia2017supersymmetric}.
 
Before the works \cite{christov2019long, manton2019forces, khare2019family, christov2019kink, christov2020kink}, some contributions to the physics of kinks with long-range tails existed already \cite{lohe1979soliton, gonzalez1989kinks, mello1998topological, gomes2012highly, khare2014successive, gani2015kink, bazeia2018analytical, belendryasova2019scattering}. 
However, in \cite{christov2019long} the authors showed that a naive initialization of the system by usual ansatz leads to wrong results, such as kink-antikink repulsion.
They showed that it is possible to achieve the correct result when the system is carefully initialized, which demands to rework some previous calculations in this regard, now that a new and more reliable method is available.
After developing a method to initialize the system correctly, the authors in \cite{christov2019long} were able to compute the interaction between the kinks and their time evolution. 
However, an accurate analytical method to estimate the force between the kinks was still lacking. One of the most used methods to estimate this force is a method developed by Manton 
\cite{manton1977force, manton1979effective}, where the force is estimated using the time derivative of the momentum density of the system. Still, in \cite{manton2019forces} the same author showed that, with the usual additive ansatz for the field, this method does not yield an accurate estimate for kinks with long-range tails and proposed a more effective method. This method consists of substituting an accelerating kink ansatz in the equations of motion and making some further approximations. This approach was proven to produce good results for a class of models with long-range tails studied in \cite{christov2019kink}. 

In the present work, we build upon the aforementioned works \cite{christov2019long, manton2019forces, christov2019kink, christov2020kink}, with some adjustments required by our model. We start with a class of $\phi^{4n}$ models that is different from the previously studied one in \cite{manton2019forces,christov2019kink} since the kink and the antikink have long-range tails on both sides, which we call double long-range. This model is relevant because it is one of the simplest models with $Z_2$ symmetry and kink solutions with double long-range tails. 
Moreover, to the best of our knowledge, this is the first investigation of the collisions of kink-antikink with long-range tails on both sides, even the ones not facing one another, and we show that the model does not exhibit resonance windows in this case.

We use the accelerating kink ansatz proposed in \cite{manton2019forces} to estimate the force between the kink and the antikink. However, it is better to solve it numerically in our case. Especially, in the case of the $\phi^{12}$ model, this is the only possibility.
Recently, the method to initialize kink-antikink collisions, in the initially static case, was extended to include the ones with moving kink and antikink applied on $\phi^{4}\phi^{2n}$ class of models \cite{christov2020kink}. However, this method includes integration of the equations of motion at each iteration step of a least-squares minimization, which is computationally costly.
In this work, we propose a new and more efficient numerical method to compute the time evolution of a kink-antikink system with a finite initial velocity. Considering the kink-antikink collisions with arbitrary initial velocities, we show that in the $\phi^{4n}$ class of models there is no bion formation in the annihilation of the kink and the antikink, except for a similar structure that appears in a small velocity interval near the critical velocity, which is ultrarelativistic. Therefore, remarkably, this model does not exhibit resonance windows originating from the lack of bion formation.

In section \ref{model}, we present the model together with the analytical results for the interaction between the kink and antikink. In section \ref{method} we present the new numerical method to initialize moving kink-antikink configurations and in section \ref{Results} we show the results of our numerical simulations. Finally, in section \ref{conclusion}, we summarize our findings.

\section{Model}
\label{model}

\subsection{Kink solution}

The model we study here consists of the following Lagrangian in $1+1$ dimensions
\begin{equation}
\mathcal{L}=\frac{1}{2}\partial_\mu\phi\partial^\mu\phi-V(\phi),
\end{equation}
where $V(\phi)$ is the scalar field potential. We consider the potentials in the form 
\begin{equation}\label{potential}
V_{2n}(\phi) = \frac{1}{2^{2n+1}} (1-\phi^2)^{2n},
\end{equation}
with $n=2$ and $n=3$, corresponding to $\phi^8$ and $\phi^{12}$ theories, respectively. Without loss of generality we chose the prefactor $1/2^{2n+1}$ for the potential to simplify some calculations. The potential can be written as \begin{equation}
V_{2n}(\phi)\equiv \frac{1}{2}\bigg(\frac{dW_{2n}}{d\phi}\bigg)^2.
\end{equation}
resulting in the following superpotentials
\begin{equation}
W_8=\frac{\phi}{4}-\frac{\phi^3}{6}+\frac{\phi^5}{20}+c_1,\quad \quad
W_{12}=\frac{\phi}{8}-\frac{\phi^3}{8}+\frac{3\phi^5}{40}-\frac{\phi^7}{56}+c_2.
\end{equation}
As the constants $c_1$ and $c_2$ are irrelevant in the rest of the calculations, we set them equal zero.

These models describe symmetry breaking with two symmetric vacua. The leading term of the expansion around these vacua is proportional to the fourth and sixth power of the scalar field, instead of the usual second power. Therefore, they have kink solutions with long-range tails interpolating between these minima and given by the BPS equation
\begin{equation}
\frac{d\phi}{dx}=\frac{dW}{d\phi}.
\end{equation}
This leads to the implicit equations for the kinks, as computed in \cite{khare2014successive}, 
\begin{equation}
\frac{2\phi}{1-\phi^2}+\log\bigg(\frac{1+\phi}{1-\phi}\bigg)=x-A,\quad\quad\frac{10\phi-6\phi^3}{(1-\phi^2)^2}+3\log\bigg(\frac{1+\phi}{1-\phi}\bigg)=2(x-A),
\label{kink-implicit}
\end{equation}
for $\phi^8$ and $\phi^{12}$ models, respectively. The local maximum of the potential is at $\phi=0$, which defines the center of the kink at $x=A$. The antikink solution, centered at $-A$, can be found setting $x\to-x$ in eq.~\ref{kink-implicit}. The masses of the kinks are given by $M=W(1)-W(-1)$ and are equal to 4/15 and 4/35 in $\phi^8$ and $\phi^{12}$ models, respectively. The asymptotic behavior of the kinks' tails was computed in \cite{khare2014successive} resulting in $1/x$ for $\phi^8$ and $1/\sqrt{2x}$ for $\phi^{12}$.

\begin{figure}[tbp]
\centering
   \includegraphics[width=\linewidth]{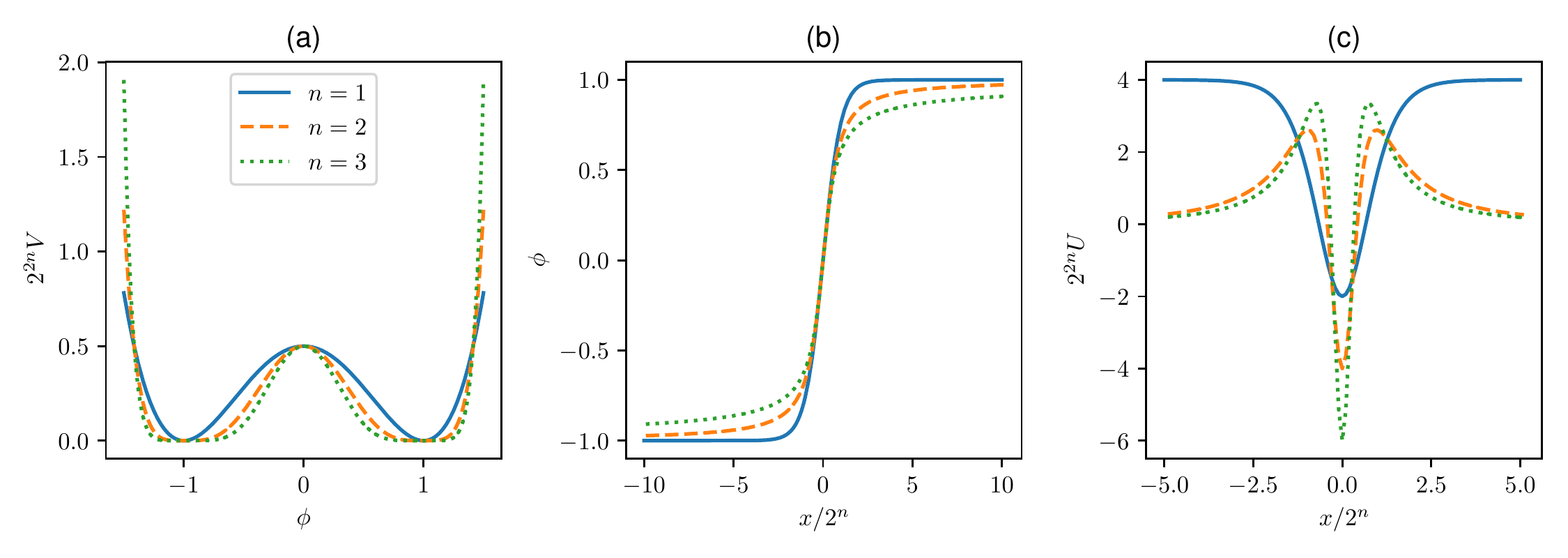}
   \caption{(a) Potential, (b) kink profile and (c) stability potential.}
   \label{kink-and-potential}
\end{figure}

The potentials and the corresponding kink solutions are shown in Fig.~\ref{kink-and-potential}. In the same figure, we have also shown the linearized potential of perturbations around the kink configuration. These potentials were computed in \cite{gomes2012highly} and have a volcano shape with vanishing asymptotic value meaning that the model does not have a shape mode. This does not necessarily mean that there will be no resonance windows if the linearized potentials have quasinormal modes \cite{dorey2018resonant, campos2020quasinormal}. However, a detailed analysis of the linearized potential is not important because we will see that this model has a strong tendency to annihilate at the collision without bion formation and, hence, it does not exhibit resonance windows. 

\subsection{Force between kink and antikink}

Recently, several analytical methods have been proposed in \cite{manton2019forces} to estimate the force between two kinks or a kink and an antikink with long-range tails. To compute the attraction between a kink and antikink with double long-range tails we adopt the most effective one among the proposed methods in the aforementioned paper. This method was shown to give accurate results for kinks with a single long-range tail \cite{christov2019kink}. 

Let us start with the configuration with well-separated kink and antikink at $A$ and $-A$, respectively. Then, we write the ansatz $\phi(x,t)=\eta(x-A(t))$ for the accelerating kink to the right of the origin and also denote the argument of $\eta$ by $X$ and the derivative with respect to it by a prime. After some approximations, the equation of motion gives \cite{manton2019forces}
\begin{equation}
\eta^\prime=\sqrt{2[V_{eff}(\eta)-aW(1)]}.
\label{int1}
\end{equation}
where $a\equiv-\ddot{A}$ is the absolute value of acceleration and $V_{eff}\equiv V+aW$ is an effective potential. In the next section, we will integrate this equation numerically to estimate the acceleration of the kinks. The rest of the calculation leads to
\cite{manton2019forces}
\begin{equation}
\label{accel}
a=\bigg[\frac{-\sqrt{\pi}\Gamma\big(\frac{n-1}{2n}\big)}{\Gamma\big(-\frac{1}{2n}\big)}\bigg]^{2n/(n-1)}\frac{A^{2n/(1-n)}}{2M},
\end{equation}
where the kink's tail and potentials were expanded up to first order. 
It is easy to see that the attraction force in the $\phi^8$ and $\phi^{12}$ models in the potential class in eq.~(\ref{potential}) has the same prefactor and power as the ones with the same long-range behavior $\phi^8$ and $\phi^{10}$, respectively, studied in \cite{christov2019kink}. 
In the derivation of the above acceleration, it is assumed that the tail of the kink that is not facing the antikink is not relevant for computing the asymptotic value of the force. This appears to be also a valid assumption in our case before the kink and antikink mostly superpose in the collision process. However, after that, this tail affects the collision behavior considerably, as we will see shortly.

If we compare the analytical expression of the acceleration to the simulations, they converge much slower compared to the result in \cite{manton2019forces,christov2019kink}. In these works, the next to the leading order term in the asymptotic tail expansion is absent, unlike our model here. Moreover, due to the long-range character of the decays, the higher-order terms become negligible relative to the first-order one much slower compared to exponential tails. Therefore, it is necessary to solve eq.~\ref{int1} numerically or include a correction in eq.~\ref{accel} due to the next to the leading order term in the tail expansion.

For the $\phi^8$ model, one can obtain the following expansion for the left tail of the kink centered at $A$, from eq.~(\ref{kink-implicit})
\begin{equation}
\label{asymptotic8}
\phi=-1+\frac{1}{A-x}+\frac{\log[2(A-x)]-\frac{1}{2}}{(A-x)^2}+\mathcal{O}\left((A-x)^{-3}\right).
\end{equation} 
If we approximate $\log[2(A-x)]\simeq\log[2A]$, the third term has a simple interpretation, which is to consider it as a shift in the point where the extrapolated tail diverges.\footnote{The idea that next to the leading-order term could be interpreted as a shift where the tail diverges was suggested in \cite{manton2019forces}.} This approximation is valid because the kink-antikink interaction is dominant in the overlapping region, where $|x|\ll A$. Thus, for $\phi^8$, we set $A\to A-\Delta A$ in eq.~(\ref{accel}) with $\Delta A=\log(2A)-\frac{1}{2}$ which has a better agreement with simulations.
For the $\phi^{12}$ model, it is hard to interpret the effect of the next to leading order term because it cannot be written as a shift in $A$. In the absence of this possibility, the system is not analytically solvable while keeping next to the leading term. Hence, we give up on an analytical estimate for this case and we solve eq.~\ref{int1} numerically instead.

It is also important to mention that the methods employed in this section are valid approximations only for small kink and antikink initial velocities. As the best match is for the static case, zero initial velocity, we will compare them to the simulations in this limit.

\section{Numerical method}
\label{method}

The numerical method employed here was first proposed in \cite{christov2019long,christov2020kink}, however, we improve the minimization of the velocity field for initially moving configurations (see below). Similarly to \cite{christov2020kink}, we consider first a traveling wave $\phi(x,t)\equiv u(x-vt)$. Substituting this expression in the equations of motion, we obtain the ODE
\begin{equation}
\label{u}
(1-v^2)u^{\prime\prime}(\xi)-V^\prime(u(\xi))=0,
\end{equation}
where $\xi=x-vt$. The velocity field of the wave is $\phi_t(x,t)\equiv f(\xi)=-vu^\prime(\xi)$. If we differentiate eq.~\ref{u} with respect to $\xi$ and multiply by $-v$ we find that the velocity field $f$ obeys another ODE given by
\begin{equation}
\label{ZM}
(1-v^2)\frac{\partial^2f}{\partial \xi^2}-V^{\prime\prime}(u(\xi))f=0.
\end{equation} 
which happens to be the Lorentz contracted zero-mode equation for a perturbation around $u(\xi)$. This is the key relation for our proposed minimization procedure for the velocity field, as we will see shortly. 

As in \cite{christov2019long,christov2020kink}, we start with the split-domain ansatz for initiating the kink-antikink collision
\begin{equation}
\phi(x,t)=(1-\Theta(x))\phi_{K}(\gamma(x+A-vt))+\Theta(x)\phi_{\bar{K}}(\gamma(x-A+vt)),
\end{equation}
where $\phi_K(x)$ ($\phi_{\bar{K}}(x)$) is the kink (antikink) solution centered at the origin and $\Theta(x)$ is the Heaviside function. To find numerical solutions, we discretize the system dividing the space into $2000$ equally spaced points, of separation $h=0.2$, in the interval $-200.0<x<200.0$, for $\phi^8$ and $h=0.4$, in the interval $-400.0<x<400.0$, for $\phi^{12}$. We also set periodic boundary conditions and approximate the second-order partial derivative with respect to $x$ by a pseudospectral matrix $D_2$ \cite{trefethen2000spectral}. After that, we have to minimize the split-domain solution to have a smooth configuration near the origin. As done in \cite{christov2019long, christov2020kink}, we find the configuration $\phi(x,t=0)\equiv u(x)$ that obeys, as close as possible, the equation for a traveling wave by minimizing the norm of the left-hand side of eq.~(\ref{u}), with the constraint that the center of the kink (antikink) should be at $A$ ($-A$).

\begin{figure}[tbp]
\centering
   \includegraphics[width=\linewidth]{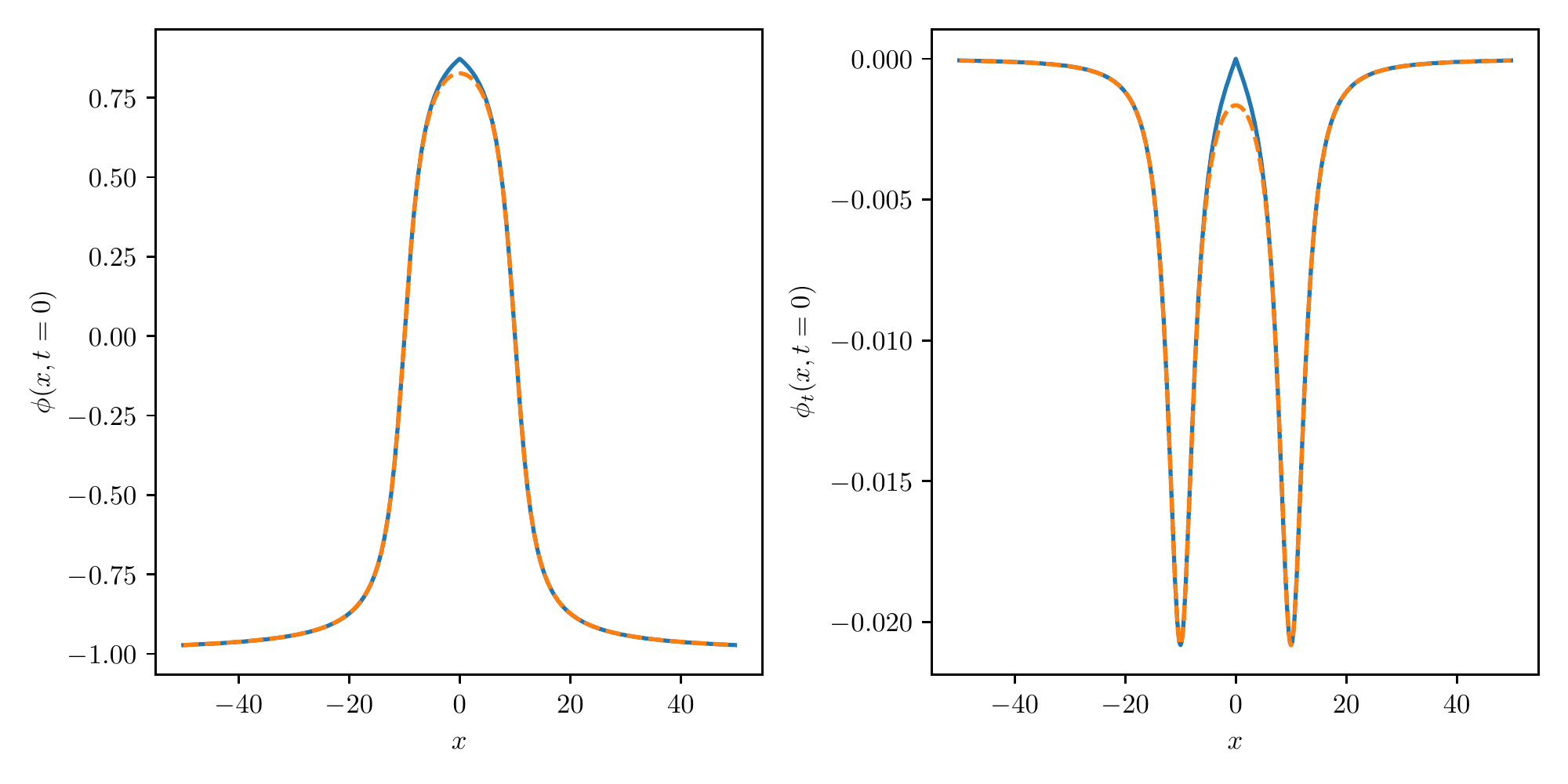}
   \caption{Initial guess (solid) and minimized solution (dashed) for the field (left) and velocity field (right). Parameters are $n=2$, $v=0.0832$ and $A=10.0$.}
   \label{minimized-solution}
\end{figure}

The next step is to write an initial guess for the velocity field. A good one is \cite{christov2020kink}
\begin{equation}
\label{phitguess}
\phi_t(x,t=0)=v\,\text{sgn(x)}u^\prime(x)
\end{equation}
This guess is similar to the split-domain ansatz in the sense that, in the domain $x<0$, it corresponds to the traveling wave solution with positive velocity and, in the domain $x>0$, to the traveling wave solution with negative velocity. This solution is correct far from the center. However, we have to minimize this initial guess to have a smooth behavior near the origin. Similarly, we demand the minimized initial solution for the time derivative of the field $\phi_t(x,t=0)$ or equivalently $f(x)$ to obey, as close as possible, the equation of a traveling wave. Therefore, we look for a function $f(x)$ that minimizes the norm of the left-hand side of eq.~(\ref{ZM}), where $u(x)$ is now the solution of the previous minimization. The minimization is subject to the constraint that $\phi_t(x,t=0$) obeys eq.~(\ref{phitguess}) in small intervals around the center of the kinks, $-A-2\leq x\leq-A+2$ and $A-2\leq x\leq A+2$ for instance. This approach to minimizing the velocity field is new and computationally more efficient than the one proposed in \cite{christov2020kink}. The initial guesses and minimized solutions for the field and velocity field are shown in Fig.~\ref{minimized-solution}.
Then, we evolve the system from the minimized initial conditions and let the kinks interact. We utilize the same set up as before for the space grid and partial derivatives with respect to $x$. Moreover, we integrate the equations of motion in time using the solve\_ivp method from the SciPy library in Python, which implements an 8th order Runge-Kutta method.

\begin{figure}[tbp]
\centering
\includegraphics[width=0.9\linewidth]{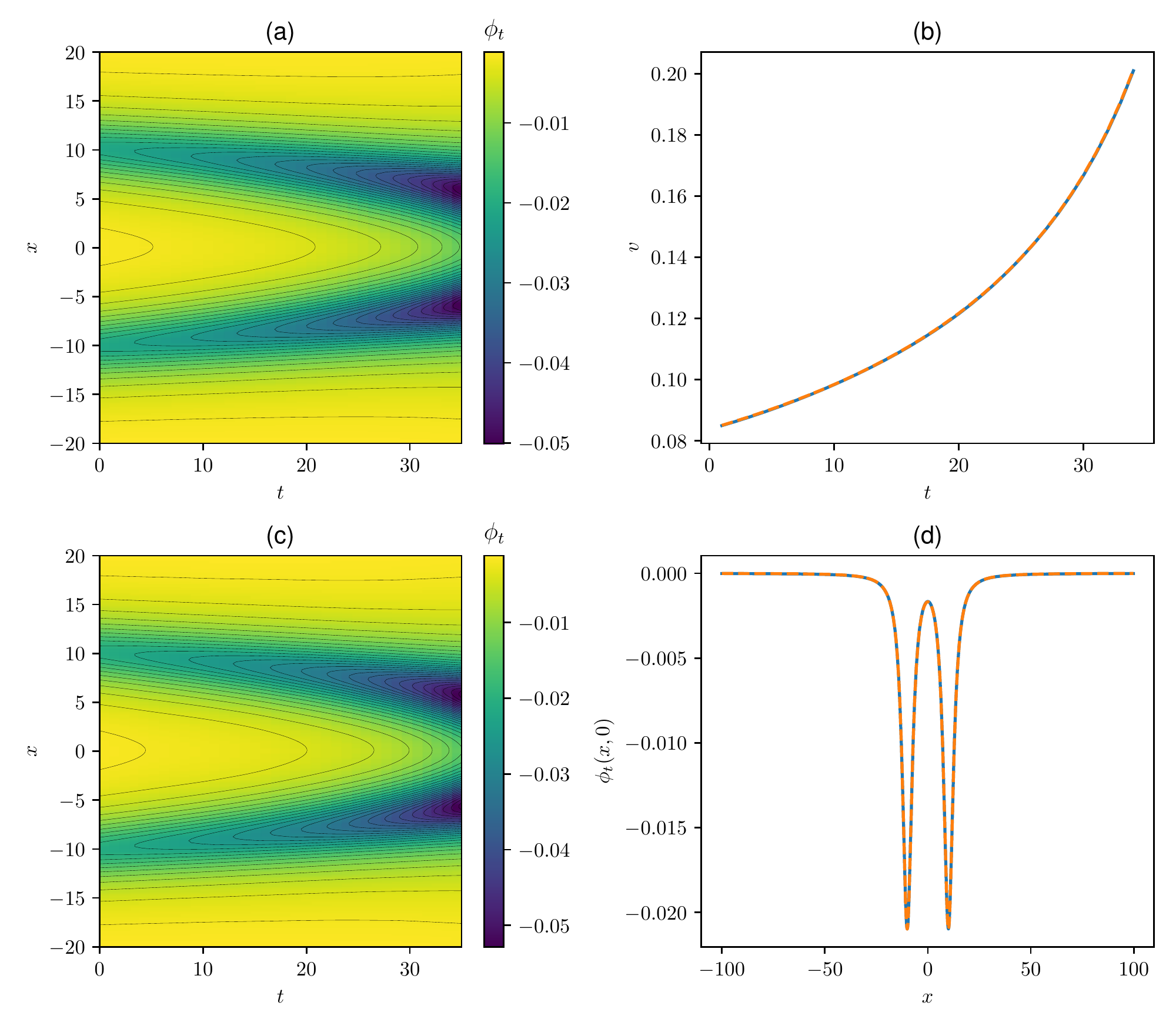}
       \caption{Plots of the evolution of the velocity field using the method with kink-antikink initially at rest (a) and our simulation with double-minimization (c). Time evolution of the velocity (b) and initial velocity field (d) for the method with kink-antikink initially at rest (dashed) and our simulation with double-minimization (solid). Parameters are $n=2$, $A=10$ and $v=0.0832$.}
       \label{phit-and-v}
\end{figure}

To test our method, we follow the same procedure as in \cite{christov2020kink} comparing the result of our simulation, with double-minimization, with a simpler initial configuration. The latter consists of starting the kink-antikink at a larger distance, $A=30$, with zero initial velocity which needs the minimization procedure only once, for the field itself. This procedure is known to produce accurate results \cite{christov2019long,christov2019kink}. In this method, we integrate the equations of motion until the center of the kink (antikink) becomes equal to or numerically as close as possible to $A=10.0$ ($A=-10.0$). As the force between the kink and antikink is attractive, they start accelerating towards each other. As they become closer, they acquire non-zero velocity where its value is dictated by the equation of motion. As a result, the final profile of the system at $A= 10$ has $v\simeq0.0832$. This is the desired kink-antikink configuration with nonzero velocity, matching the same input used in our simulation with double-minimization. At this point,  we can continue the evolution from this configuration as a reference evolution for a kink-antikink configuration.  In Fig.~\ref{phit-and-v} we plot the evolution of the velocity field, the kink velocity, and the initial velocity field utilizing both methods. As one can see, the velocity matches and evolves smoothly in the two methods. Moreover, the initial velocity field, as well as its evolution in our simulation, also has an excellent agreement with the method where the kink and antikink are initially static. This shows that our minimization procedure is correctly initializing the system.

\section{Results}
\label{Results}

\begin{figure}[tbp]
\centering
   \includegraphics[width=\linewidth]{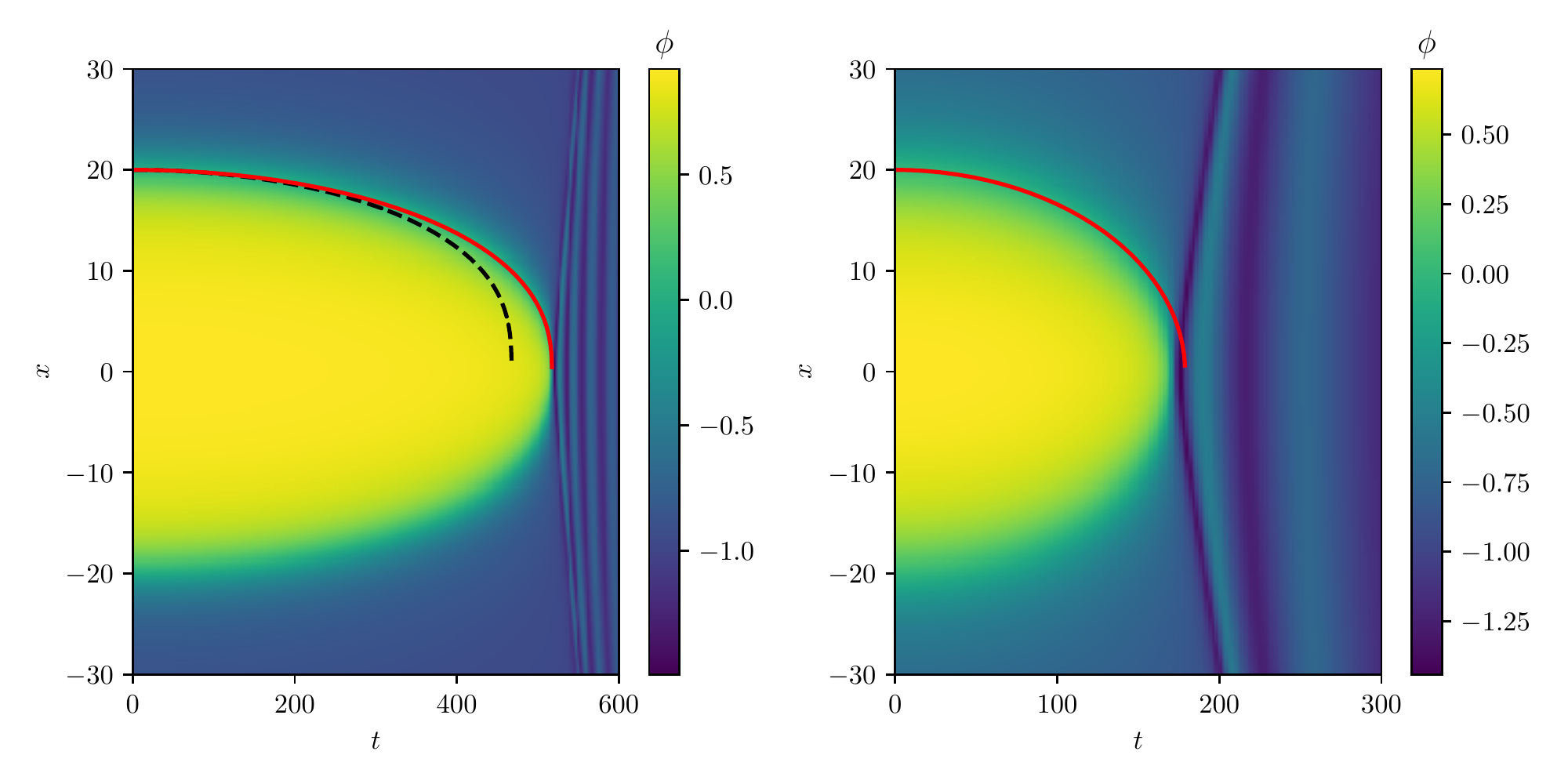}
   \caption{Field evolution for $\phi^8$ (left) and $\phi^{12}$ (right) models. The initial condition is the minimized field solution with $A=20.0$. The superimposed curves are the integration of the eq.~(\ref{accel}) including the effect of the shift $\Delta A$ (dashed) and the numerical solution of eq.~(\ref{int1}) (solid).}
   \label{field-evolution}
\end{figure}

We start by integrating the equations of motion when the kink and antikink are initially at rest. The method to initialize this system was proposed in \cite{christov2019long} and is a special case of the method described in Sec.~\ref{method} with the vanishing velocity field. The result is shown in Fig.~\ref{field-evolution}. As the higher-order terms in the asymptotic expansion of the tails in our model are important, we must solve eq.~(\ref{int1}) with higher-order terms to estimate the acceleration. In this case, the best way is to solve it numerically. Hence, we can keep terms of all orders and solve them without further approximations. The integration of this numerical solution is shown with solid curves for both models in Fig.~\ref{field-evolution} and evidently is in good agreement with the simulations, which serves as a consistency check. For the $\phi^8$ model, we also superimpose on the image the numerical integration for the theoretical result in eq.~(\ref{accel}) with the prescription $A\to A-\Delta A$ which works relatively well, although with noticeable deviation. In the $\phi^{12}$ model, we cannot make the same approximations and, thus, we only compute the force numerically. The force estimate for both models, especially in $\phi^{12}$, becomes more accurate as $A$ is increased. The same is true for the asymptotic expression of eq.~(\ref{accel}), but it converges more slowly to the simulated values as $A$ is increased. 

To compute the numerical acceleration, which comes from eq.~(\ref{int1}) without further approximations, we match the accelerating kink profile as close as possible to the static kink profile. The approach we adopt is to match the center of both kinks. This is done integrating eq.~(\ref{int1}) for some value of $a$ for x between $0$ and $A$ or equivalently $X$ between $-A$ and $0$. The integration starts from an initial value $\eta(-A)$ such that $\eta^\prime(-A)=0$ in eq.~(\ref{int1}). This way it becomes a root-finding problem which consists of finding the value of the variable $a$ for which the field $\eta(0)=0$. We solve this problem with the bisection method, which converges linearly. The final value of $a$ is the estimate of the acceleration of the system. To find the solid theoretical curve shown in Fig.~\ref{field-evolution}, we integrate the result and find the position as a function of time, computing the acceleration at each time step. Notice that, in this procedure, we only match the tail that is facing the opposing kink to get an accurate estimate of the acceleration. Therefore, the other tail is not relevant to the interaction when the kinks are far away.

One should notice that after the collision the kink and antikink annihilate directly into radiation, instead of forming a bion first which is in clear contrast with the ones in $\phi^4$ theory or even the kink solutions with one long-range tail. We argue that this happens because the potential near both vacua is nearly flat (see Fig.~\ref{kink-and-potential}). This, in turn, occurs because the leading term in the expansion around both of them is of a higher order. To understand better why the model behaves differently in our case, let us discuss what happens at a kink-antikink annihilation in the three scenarios: kinks with exponentially decaying tails, kinks with one power-law tail, and kinks with power-law tails on both sides. 

First, for kinks with exponentially decaying tails, take $\phi^4$ for example, we start with an antikink on the left and a kink on the right. Eventually, after the collision, the field configuration will cross the barrier at $\phi=0$, and the configuration will be near $\phi=1$ all over space. Notice that the vacuum at $\phi=1$ is a regular one, with quadratic leading order term in the expansion. After the field reaches this configuration, it keeps decreasing until there is a rebound and the field crosses the barrier again, recreating the kink-antikink pair. For kinks with one power-law tail, take the $\phi^{4+2n}$ model in \cite{christov2019long,christov2019kink} for example, the vacuum at $\phi=0$ has a higher-order expansion, while at $\phi=1$ it is quadratic. Suppose that there is an antikink in the $(1,0)$ sector on the left and a kink in the $(0,1)$ sector on the right. In this configuration, the long-range tails are facing each other. After the collision, the field will also eventually cross the potential barrier, which is somewhere in the range $0<\phi<1$, and will be near $\phi=1$ all over space. In this case, the system behaves similar to $\phi^4$ theory because the vacuum at $\phi=1$ is also quadratic to leading order. That is, the field will continue decreasing until there is a rebound and the kink-antikink pair is created again. 

Now, consider a model with kink solutions that has power-law tails on both sides, take our model here as an example, also starting with an antikink on the left and a kink on the right.
Again the field will eventually cross the potential barrier at $\phi=0$ after the collision and will be near $\phi=1$ all over space. However, this time the potential around $\phi=1$ is nearly flat and the energy density will spread quickly, almost like a free field. Therefore, there will not be enough energy near $x=0$ to cross the barrier at $\phi=0$ after the rebound and create another kink-antikink pair. This means that the behavior of double long-range tail models is different from the behavior of the ones with exponential tails or even a single long-range tail 
\cite{christov2019long}. This is because there is annihilation directly into radiation without the bion formation for small initial velocities. Moreover, we will see that, even for large initial velocities, there is still no long-lived bion formation after the kink and the antikink annihilate. This is in sharp contrast with previously studied models.

\begin{figure}[tbp]
\centering
   \includegraphics[width=\linewidth]{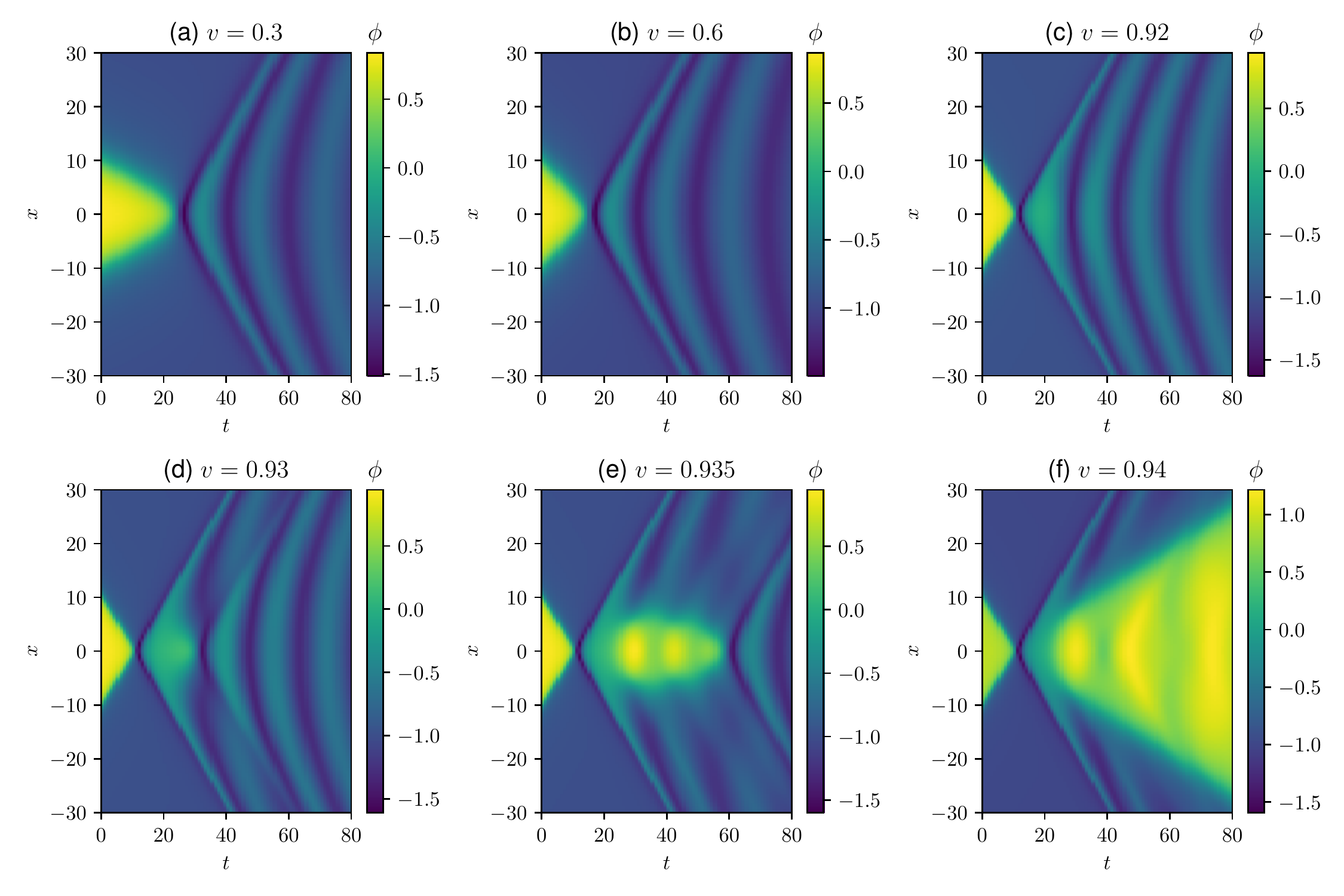}
   \caption{Collisions for the $\phi^8$ model with different values of $v$ and initial position $A=10.0$.}
   \label{collisions8}
\end{figure}

Now, let us study the kink-antikink collisions for nonzero initial velocities. Simulation results for various initial velocities are shown in Fig.~\ref{collisions8} considering $n=2$, $\phi^8$ model. As the velocity becomes nonzero we still observe kink-antikink annihilation directly into radiation, as observed in the $v=0$ case. This occurs until we are very close to the critical velocity, which is ultrarelativistic in this model. This shows that when the "backtail", the tail not facing the other kink, is long-range, the collision behavior is drastically altered. As argued in \cite{campbell1983resonance}, it is not trivial beforehand whether a kink-antikink collision for a given model should annihilate directly into radiation or form a bion. Our heuristic argument we gave above is a qualitative justification for the suppression of the bion formation in the class of models with kinks with long-range backtails, shedding light on the subject.

For $v=0.93$ and $v=0.935$, the behavior starts to change. We see that the kink and antikink reflect, but the velocity is not large enough for them to separate. Thus, they collide a second time and annihilate. This resembles a bion formation but is not exactly the same because this bound state is always annihilated at the second collision. The velocity after the first collision is much smaller and annihilation occurs at the second bounce similar to collisions with small initial velocities. This means that this system does not exhibit resonance windows because too much radiation is generated at bounces. The absence of resonance windows is another property of systems with long-range "backtails". Finally, at the critical velocity of $v_c=0.94$ the kink and antikink reflect. It is important to emphasize that this value of $v_c$ was obtained for initial position $A=10$ and, due to the long-range interaction between the kinks, the value of $v_c$ can be slightly different for different values of $A$. 

\begin{figure}[tbp]
\centering
   \includegraphics[width=\linewidth]{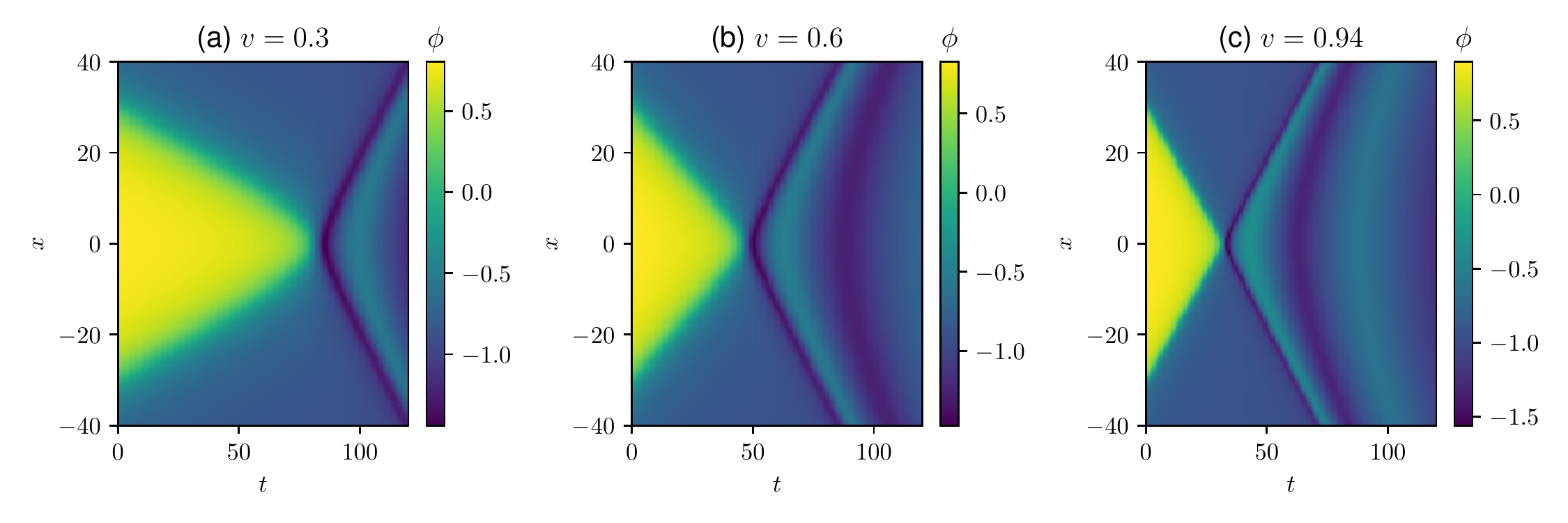}
   \caption{Collisions for the $\phi^{12}$ model with different values of $v$ and initial position $A=30.0$.}
   \label{collisions12}
\end{figure}

The result of the simulations for the collisions in the $\phi^{12}$ model is shown in Fig.~\ref{collisions12}. For this model, the tail is even more long-range and the overlap between the kink and the antikink is larger. To minimize the overlap and have a more accurate minimization of the velocity field we increased the separation and chose $A=30$ in this case. Again, as expected, there appears just annihilation without the bion formation below the critical velocity.  As the kink and antikink in this model have fatter tails, the critical velocity is even higher (close to 1), compared to the $\phi^8$ model, which makes it more difficult to find numerically.  This is because the minimization of the velocity field is difficult to converge when $v$ approaches $1$. Nevertheless, we expect that at some point the kink-antikink energy will be high enough such that they will separate after bouncing. 

\section{Conclusion}
\label{conclusion}

In this work, we have investigated a class of $\phi^{4n}$ models, with $n=2$ and $n=3$ specifically, where the potential has $Z_2$ symmetry with two minima and the kink and antikink have long-range tails on both sides. 
Due to the long-range character of the tails, we have observed that the interaction decays as a power-law with the kinks separation, instead of the usual exponential decay, making them interact strongly. We showed that the method developed by Manton based on the accelerating kink ansatz gives convincing results for the acceleration of the kink in our model when solved numerically before the kink and antikink superpose to great extent. We solve it numerically because the convergence to the asymptotic value is slow due to the long-range character of the higher-order terms in the tail expansion. However, we were able to give an analytical estimate of the correction in the acceleration expression for the $\phi^8$ model interpreting next to the leading-order term in the tail expansion as a shift in the point where the leading-order term diverges and the result is compatible with the numerical simulations.

We have integrated the equations of motion for an initial configuration with a kink and an antikink with a finite initial velocity. As the kinks have long-range tails, naive use of a typical ansatz to initialize the system, such as the additive ansatz, does not lead to an accurate description of the system and specialized methods should be used. Thus, we have adopted the method developed by Christov et al. \cite{christov2019long, christov2020kink} to minimize the split-domain ansatz to a configuration that obeys the equations of motion of a moving kink as closely as possible and we developed a more computationally efficient method to initialize the velocity field. Our method consists of finding an equation for the velocity field of a traveling wave and looking for an initial velocity field of a kink-antikink configuration that obeys this equation as close as possible.

The results of the numerical simulations of the equations of motion show that the kink and antikink annihilate directly into radiation instead of forming a bion first, with the exception that a similar structure appears in a small velocity interval near the critical velocity. This is in stark contrast with the models with kinks that have exponentially decaying tails or even the ones with a single long-range tail. We have argued that this occurs because the potential is nearly flat near both minima and, therefore, the energy spreads rapidly during the collision. This way, the system does not have enough energy near the center of the collision to create yet another kink-antikink pair. 
Moreover, we found the ultrarelativistic critical velocity $v_c=0.94$ for the $\phi^8$ model when the initial separation is $A=10.0$, as well as short-lived bound kink-antikink states for velocities close to this value. Interestingly, The tendency to annihilate into radiation forces the kink and the antikink to annihilate at the slower second bounce and, therefore, prohibits resonance windows in this system. In fact, the absence of resonance windows made it less relevant to present a detailed analysis of the stability equation in this paper.

The study of kinks with long-range tails has not been explored thoroughly enough so far, and there remain many possibilities for future work. There are many classes of scalar field theories in $(1+1)$ dimensions with polynomial potentials that exhibit long-range tails, as summarized in 
\cite{khare2014successive, khare2019family}, and some could be interesting to explore and could possibly contain new physical phenomena. We believe that our method for initializing the kinks with nonzero velocity is a highly computationally efficient method to initialize these systems. It would also be interesting to see if similar constructions could be used to initialize a solitonic system with power-law asymptotics in higher dimensions. One more direction we plan to explore is the interaction of kinks with a single long-range tail when the long-range tails do not face each other. From our current analysis, we expect that the long-range tails would not affect the force between them for large separations but would change the bion formation after the collision and the existence of resonance windows. However, a more detailed analysis is needed to verify this intuitive guess.

\section*{Acknowledgments}

We acknowledge financial support from the Brazilian agencies CAPES and CNPq. AM also thanks financial support from Universidade Federal de Pernambuco Edital Qualis A.

\end{document}